\documentclass[runningheads]{llncs}
\usepackage[T1]{fontenc}
\usepackage{graphicx}
\usepackage{amsmath}
\usepackage{amssymb}
\usepackage{caption}
\usepackage{booktabs}
\usepackage{array}
\usepackage{hyperref}
\usepackage{float}
\usepackage{indentfirst}
\usepackage{placeins}
\begin{document}

\title{Probabilistic AVL Trees (p-AVL): Relaxing Deterministic Balancing}

\author{Hayagriv Desikan}
\institute {Indian Institute of Technology,Jodhpur}

\maketitle

% -------------------------------------------------------
\begin{abstract}
This paper studies the empirical behaviour of the p-AVL tree, a
probabilistic variant of the AVL tree  in which each imbalance is
repaired with probability $p$. This gives an exact continuous
interpolation from $p = 0$, which recovers the BST endpoint, to $p = 1$,
which recovers the standard AVL tree. Across random-order insertion
experiments, we track rotations per node, total imbalance events,
average depth, average height, and a global imbalance statistic $\sigma$.
The main empirical result is that even small nonzero $p$ already causes a
strong structural change. The goal here is empirical rather than fully
theoretical: to document the behaviour of the p-AVL family clearly and
identify the main patterns.
\end{abstract}

% -------------------------------------------------------
\section*{Introduction}

We study the empirical properties of the p-AVL tree. A p-AVL tree is a probabilistic relaxation of an AVL tree[1] in which each triggered repair is applied with probability $p$ rather than deterministically. The recursive unwinding process is otherwise unchanged.

This creates a continuous interpolation from $p = 0$ to $p = 1$, with $p$ treated as a continuous parameter in $[0,1]$. The BST is recovered at $p = 0$ and the regular AVL tree at $p = 1$. So, instead of a single trade-off, we have a regime parametrised by $p$.

The main quantities examined in this paper are rotations/$N$, the total imbalance count encountered during tree formation, average depth, average height across runs, and $\sigma$, a weighted measure of global
structural imbalance.

% -------------------------------------------------------
\section{Rotations per Node}

We first examine rotations/$N$ as a function of $p$ across $N$ from 8{,}000 to
512{,}000, using logarithmic spacing in $p$ with especially high resolution
in the very small-$p$ regime. The main observation is a near-collapse of
the curves across $N$. A simple base model, taken from prior fitting work,
is $f(p) = 0.67(1 - e^{-5.72p})$. The fit is already strong, but the
residual is structured rather than random, which motivates the
second-stage residual analysis below. The base model statistics are:

\begin{itemize}
    \item MSE: $3.553 \times 10^{-4}$
    \item RSE: $1.885 \times 10^{-2}$
    \item Pearson: $0.993523$
    \item Residual variance: $3.008 \times 10^{-4}$
    \item Residual std: $1.734 \times 10^{-2}$
\end{itemize}

The following plot shows rotations/$N$ versus $p$ for various $N$ from 8{,}000
to 512{,}000. The $p$-grid uses logarithmic spacing, with especially dense
sampling from $10^{-6}$ to $10^{-3}$. Notably, the curves exhibit a near-collapse
across $N$.

\begin{figure}[ht]
    \centering
    \includegraphics[width=\textwidth]{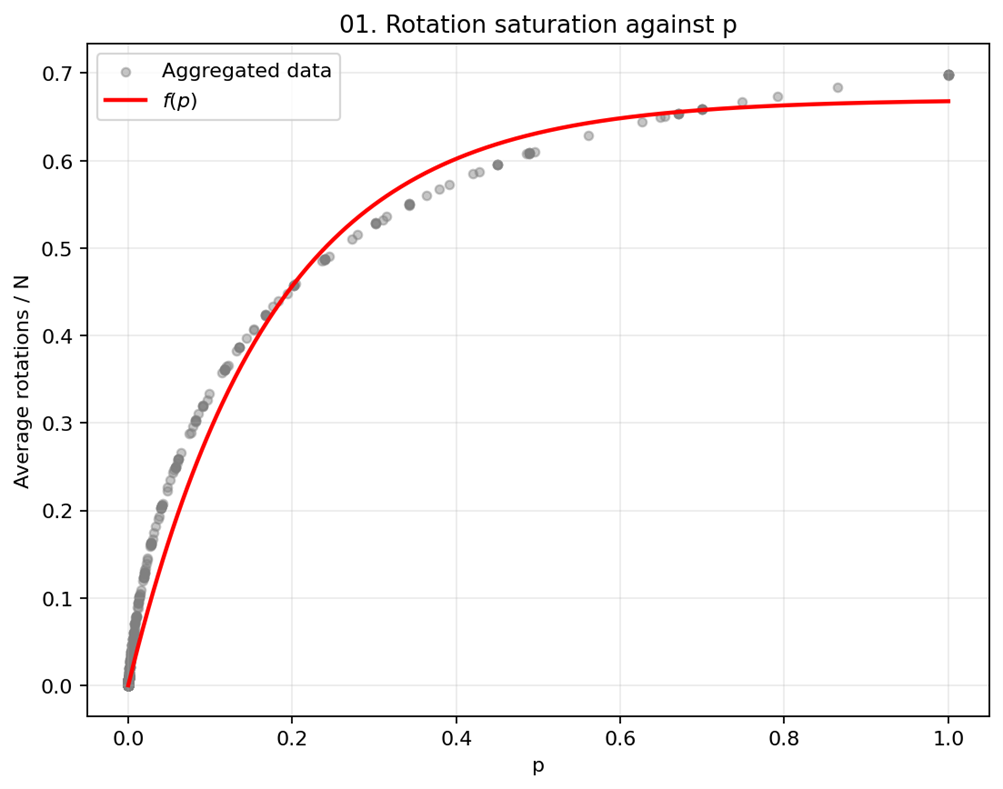}
    \caption{Rotations per node (rotations/$N$) versus $p$ for $N$ ranging
    from 8{,}000 to 512{,}000. Logarithmic spacing in $p$ is used, with dense
    sampling in the small-$p$ regime. The curves exhibit near-collapse across $N$.}
    \label{fig:rot_per_node}
\end{figure}

After extensive curve fitting, the residual was modelled empirically as
\begin{equation}
    R(p) = k \cdot \varphi(f) \cdot \left[\varphi(f) - a\right] \cdot \left[\varphi(f) - b\right]
\end{equation}
where $\varphi$ is a nonlinear warp of $f(p)$:
\begin{equation}
   \varphi(f) = \frac{f(p) + d_1 \cdot f(p)^2}{1 + a_1 \cdot f(p) + a_2 \cdot f(p)^2 + a_3 \cdot f(p)^3}
\end{equation}

The exact parameter variance across $N$ was not analysed due to the
structured and visual $N$-invariant nature of the residual, coupled with
the fact that the $y$-axis terms were already in the order of $10^{-2}$ terms.

\begin{figure}[ht]
    \centering
    \includegraphics[width=0.75\textwidth]{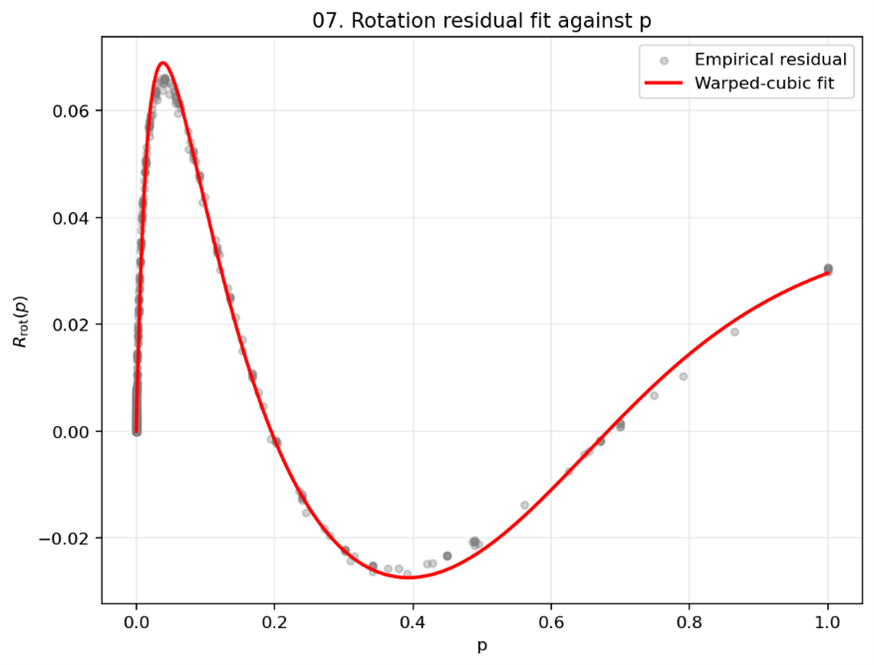}
    \caption{Residual $R(p) = \text{rot}/N - f(p)$ after subtracting the base
    exponential model, showing structured cubic-like behaviour before the
    nonlinear warp is applied.}
    \label{fig:residual_rot}
\end{figure}

The parameters are as follows:
\begin{itemize}
    \item $a_1$: 2.71747654
    \item $a_2$: $-10.00000000$
    \item $a_3$: 5.69933709
    \item $d_1$: $-1.45418721$
    \item $a$: 0.21760284
    \item $b$: 0.34390392
    \item $k$: 23.00000000
\end{itemize}

This parameter ($k$) is partly degenerate with $a_2$, so the more stable
quantities to interpret are the raw zeroes $a$ and $b$ together with
$d_1$, rather than every optimiser output taken literally.

Combining the base term and residual gives $\text{rot}/N = f(p) + R(p)$.
The resulting empirical model gives the following overall statistics:

\noindent\textbf{Combined rotation model statistics:}
\begin{itemize}
    \item MSE: $3.396 \times 10^{-6}$
    \item RSE: $1.849 \times 10^{-3}$
\end{itemize}

The residual form was found by inspecting $\text{rot}/N - f(p)$ and then
searching for a change of variables that turns the tilted, compressed
structure into something close to a cubic. The point is not that this is
the unique possible model, but that it provides a compact and useful
empirical fit.

A similar residual pattern appears again in the next section, which is
one of the main recurring empirical observations of the paper. The below
plots demonstrate how the non-linear warp turns the above
stretched/distorted cubic into a normal cubic.

\begin{figure}[ht]
    \centering
    \includegraphics[width=0.48\textwidth]{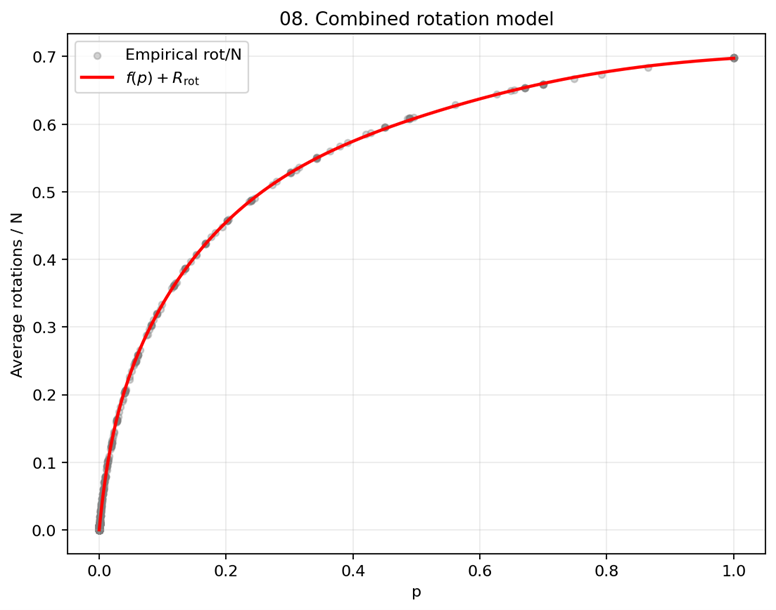}\hfill
    \includegraphics[width=0.48\textwidth]{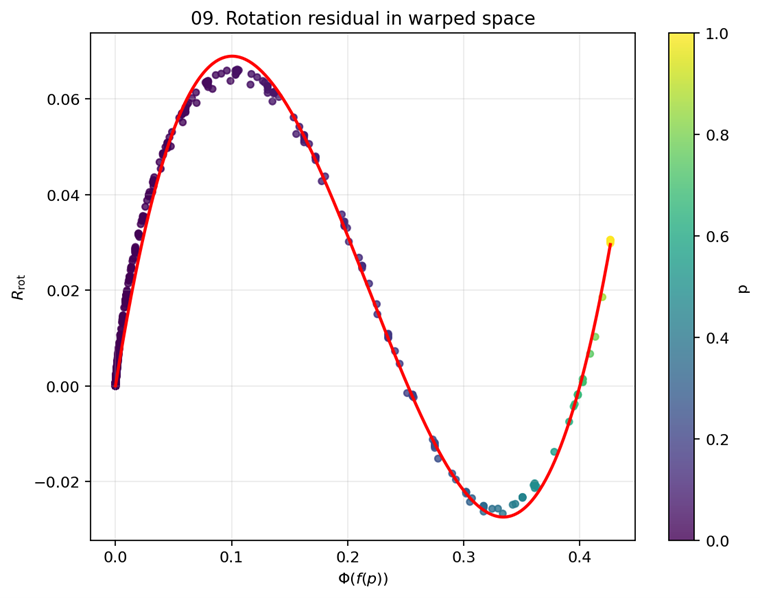}
    \caption{Left: the combined model of $\mathrm{Rot}/N = f(p) + R(p)$, indicating a high-quality visual fit. 
Right: the fit of the residual with the non linear warp,indicating how we recover a regular cubic from a warped one in p-space into the warp}
    \label{fig:warp_comparison}
\end{figure}

The table below shows the fitted zeroes $a$ and $b$ across $N$. Their near
invariance is one of the clearest signs that the residual structure is
stable across system size.

\begin{table}[ht]
    \centering
    \caption{Fitted zeroes $a$ and $b$ of the residual model across different
    values of $N$. Their near-invariance confirms that the residual structure
    is stable across system size.}
    \label{tab:zeroes_ab}
    \begin{tabular}{@{}lcc@{}}
        \toprule
        \textbf{N} & \textbf{a} & \textbf{b} \\
        \midrule
        8{,}000   & $2.249629 \times 10^{-1}$ & $3.599847 \times 10^{-1}$ \\
        16{,}000  & $2.266764 \times 10^{-1}$ & $3.608127 \times 10^{-1}$ \\
        32{,}000  & $2.279989 \times 10^{-1}$ & $3.606134 \times 10^{-1}$ \\
        64{,}000  & $2.288234 \times 10^{-1}$ & $3.606846 \times 10^{-1}$ \\
        128{,}000 & $2.293406 \times 10^{-1}$ & $3.606926 \times 10^{-1}$ \\
        256{,}000 & $2.295806 \times 10^{-1}$ & $3.609341 \times 10^{-1}$ \\
        512{,}000 & $2.295457 \times 10^{-1}$ & $3.608161 \times 10^{-1}$ \\
        \bottomrule
    \end{tabular}
\end{table}

The corresponding values of $p^*_a$ and $p^*_b$ are obtained by solving
$\varphi(f(p)) = a$ and $\varphi(f(p)) = b$. They also vary only weakly with $N$.

\begin{table}[ht]
    \centering
    \caption{Crossing points $p^*_a$ and $p^*_b$ in $p$-space, derived from
    the fitted zeroes $a$ and $b$. Both values show only weak dependence on $N$.}
    \label{tab:crossing_points}
    \begin{tabular}{@{}lcc@{}}
        \toprule
        \textbf{N} & \textbf{$p^*_a$} & \textbf{$p^*_b$} \\
        \midrule
        8{,}000   & 0.190746 & 0.697421 \\
        16{,}000  & 0.194355 & 0.692111 \\
        32{,}000  & 0.193882 & 0.691976 \\
        64{,}000  & 0.195223 & 0.686445 \\
        128{,}000 & 0.196438 & 0.686519 \\
        256{,}000 & 0.196561 & 0.686103 \\
        \bottomrule
    \end{tabular}
\end{table}

\noindent Variance across $N$ of $p^*_a = 4.37 \times 10^{-6}$, variance of
$p^*_b = 1.78 \times 10^{-5}$.

Across $N = 8{,}000$ to $512{,}000$, the crossing points $p_a$ and $p_b$ show only
weak dependence on $N$. These are zeroes in $\varphi(f(p))$ space mapped back
into $p$-space, so they should be interpreted as features of the chosen
empirical model rather than literal intrinsic zeroes of the p-AVL tree.
That said, their weak $N$-dependence is still notable and helps explain
why the residual structure looks visually stable.

One may argue that the choice of 0.67 is partly empirical, since in
simplifying the model to fix the constant $A(1-e^{-bp})$, where $A$ is the
effective saturation value for $\mathbb{E}[\text{rotations}/N]$ in an AVL tree
under random insertion, standard theoretical analysis does not admit an
exact closed-form value for $A$. Choosing a value closer to 0.697 or 0.7
could instead lead to the second zero being pushed further toward the
endpoint rather than being eliminated entirely, which would not only
complicate the analysis, but would also make the resulting structure less
clean compared to the nearly evenly placed values of $p^*_a$ and $p^*_b$
obtained in the chosen model. This objection does, however, confirm an
important point, namely that the values of $p^*_a$ and $p^*_b$ are
model-dependent and are not true zeroes of the p-AVL tree itself, but
rather depend on the chosen mean-field regime.

A direct fit for rot/$N$ was not pursued, since the raw exponential model
already admits a high degree of accuracy, with high correlation and good
values of RSE and MSE. Along with this, the residual, namely
$\text{rot}/N - f(p)$, when compared with the chosen $f(p)$, has a cosine
similarity of 0.182, indicating that it is not simply a linear extension
of the original $f(p)$, but instead retains a largely distinct structure,
as already demonstrated by prior plots.

% -------------------------------------------------------
\section{Imbalance Events}

A natural first guess is that total rotation count should scale with the
number of imbalance events,consistent with classical observations on update behaviour in binary search trees [3],
modulated by the repair probability $p$.
Ignoring higher-order effects, this suggests that
$p \times \text{imbalances\_total}$ should be approximately linear in the raw
rotation count. Empirically, this relation holds very strongly.

A better empirical interaction model is as follows:
$\text{imbalances} \times p = m \times \text{rotations}_{\text{raw}} + \lambda \times N \times p$.
We observe a very rapid collapse in the raw number of imbalances
encountered, especially at small $p$. As $p$ moves from 0 to 0.1, a large
fraction of imbalance events is already eliminated across $N$.

\begin{figure}[ht]
    \centering
    \includegraphics[width=0.7\textwidth]{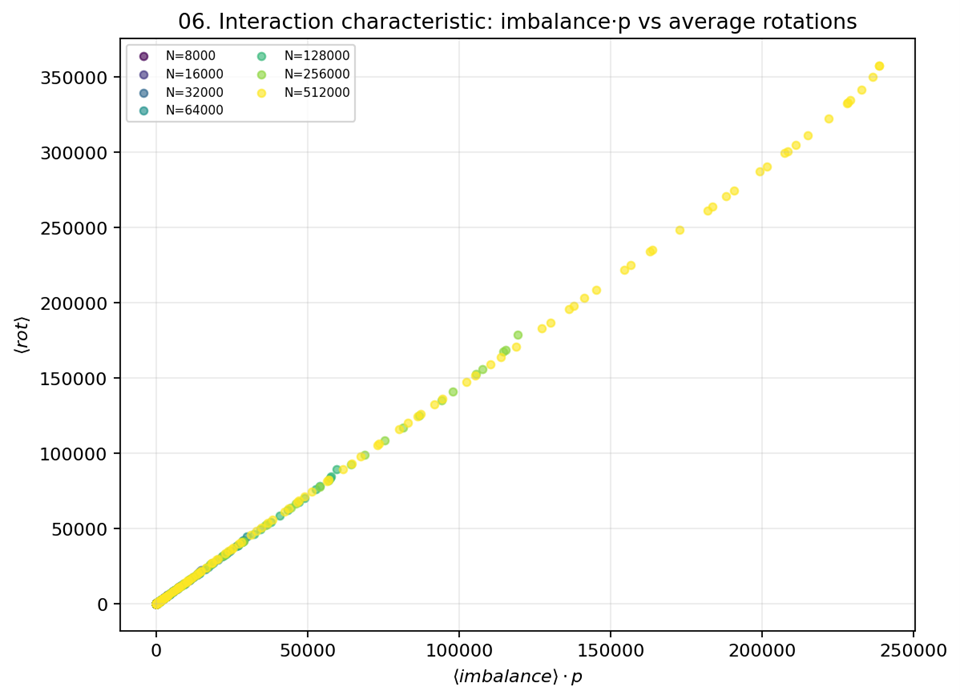}
    \caption{Total imbalance count scaled with p for various N compared with raw rotation count}
    \label{fig:imbalance_events}
\end{figure}

Further fitting used the model:
\[
\text{imbalances} \times p = m \times \text{rotations}_{\text{raw}} + \lambda \times N \times p
\]

\noindent\textbf{Interaction model drift:}
\begin{itemize}
    \item $m$: $0.703972\ (\pm 4.772 \times 10^{-5})$
    \item $\lambda$: $-0.020980\ (\pm 8.233 \times 10^{-5})$
\end{itemize}

\noindent\textbf{Interaction model statistics:}
\begin{itemize}
    \item MSE: $4.998 \times 10^{4}$
    \item RSE: $2.238 \times 10^{2}$
    \item Pearson: $0.999979$
    \item Residual variance: $4.899 \times 10^{4}$
    \item Residual std: $2.213 \times 10^{2}$
\end{itemize}

The residual variance and residual standard deviation look large in
absolute terms because the raw imbalance and rotation values are
themselves large. For that reason, the plots are more informative here
than the raw scale of those residual summary statistics alone.

This figure matters because the residual for imbalances does not have to
resemble the residual for rotations/$N$. Empirically, however, it again
shows a stretched, warped cubic-like form after a nonlinear warp. That
recurrence is suggestive, even though it should still be treated as an
empirical pattern rather than a proved mechanism.

\begin{figure}[ht]
    \centering
    \includegraphics[width=0.7\textwidth]{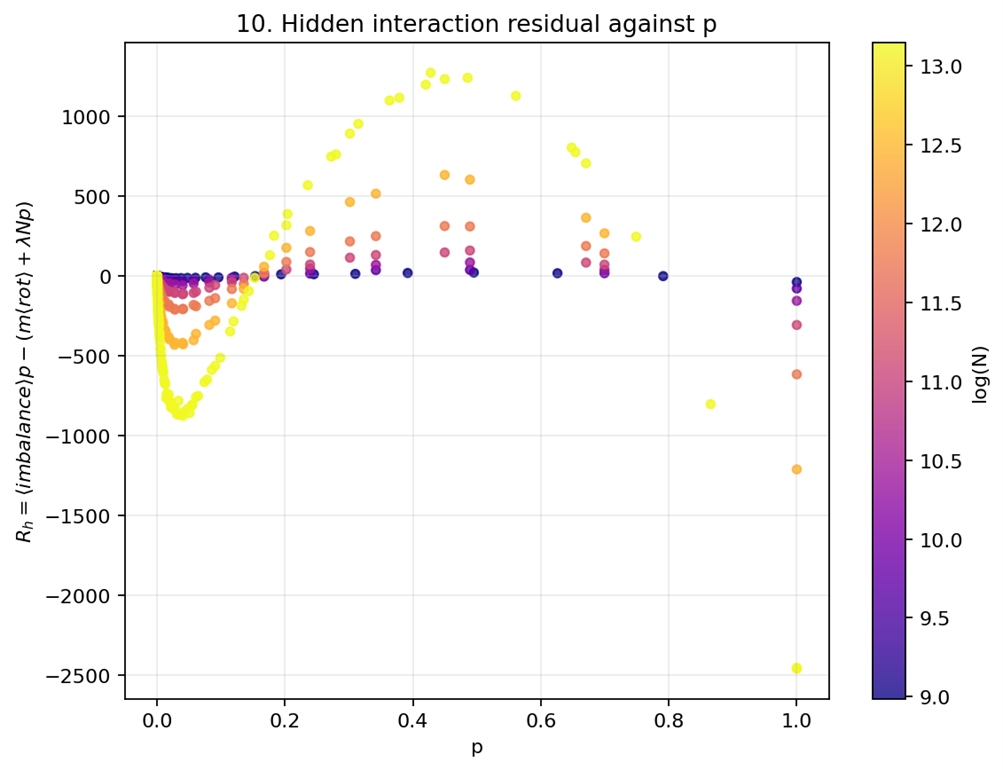}
    \caption{Residual of the imbalance interaction model in original $p$-space.
    The structure resembles a stretched, warped cubic, analogous to the residual
    seen in the rotations/$N$ model.}
    \label{fig:imbalance_residual1}
\end{figure}

\begin{figure}[ht]
    \centering
    \includegraphics[width=0.7\textwidth]{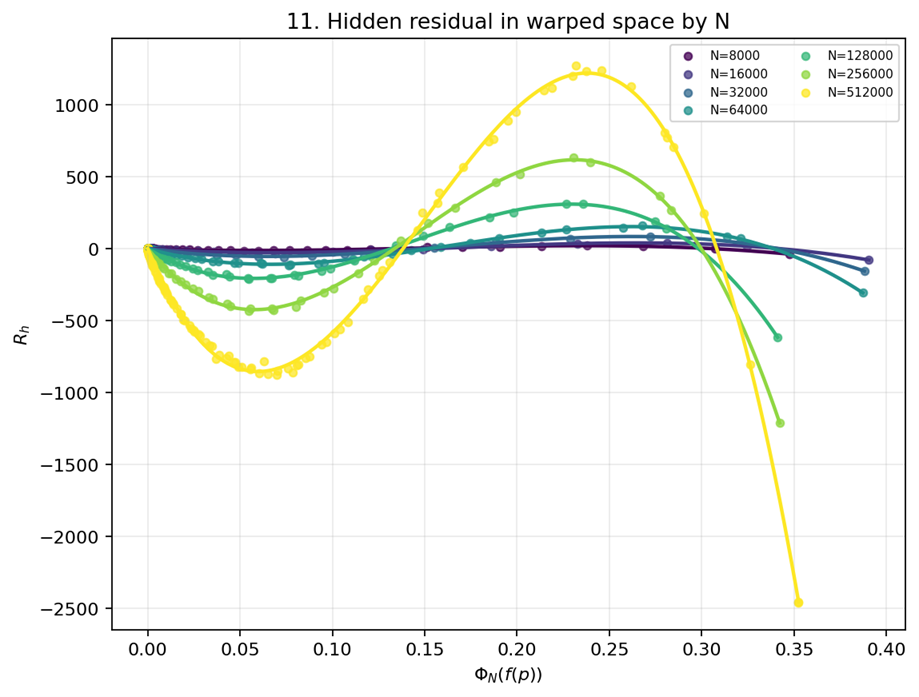}
    \caption{Imbalance residual after applying the nonlinear warp $\varphi$,
    revealing a cleaner cubic structure. The recurrence of this form across
    sections is a notable empirical pattern.}
    \label{fig:imbalance_residual2}
\end{figure}

The exact parameter variance across $N$ was not analysed in detail because
the residual appears visually stable across $N$. The large residual
variance and standard deviation mainly reflect the large raw scale of
the imbalance and rotation counts.

Here,
\begin{equation}
    R(N,p) = k(N) \cdot \varphi(f) \cdot \left[\varphi(f) - a\right]
              \cdot \left[\varphi(f) - b\right]
\end{equation}

The residual fit for imbalances $\times\, p$ is accurate enough for our
purposes. As in the previous section, parameter degeneracy is present,
so the more useful quantities to compare across $N$ are the identifiable
features rather than every raw optimiser output. $k(N)$ follows the
relationship shown below.

\begin{figure}[ht]
    \centering
    \includegraphics[width=0.7\textwidth]{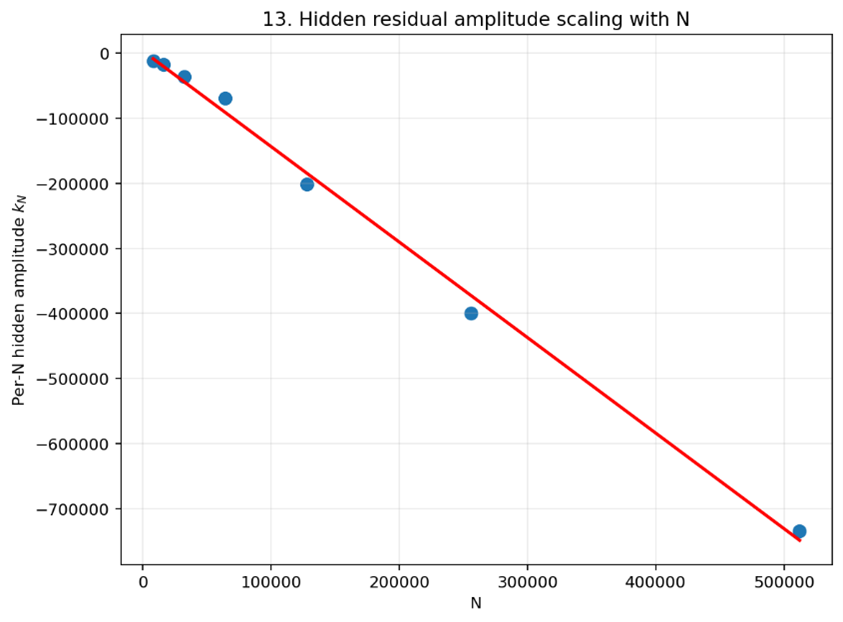}
    \caption{Scaling of the amplitude $k(N)$ in the imbalance residual model
    as a function of $N$. The scalar $k$ grows with $N$, setting the overall
    magnitude of the residual.}
    \label{fig:k_scaling}
\end{figure}

The residual for imbalances $\times\, p$ need not resemble the residual for
rotations/$N$. Empirically, however, it again shows a stretched, warped
cubic-like form after a nonlinear warp. The main difference is that its
magnitude scales with $N$, unlike rotations/$N$, which nearly collapses.

\begin{table}[H]
    \centering
    \caption{Fitted parameters $a$, $b$, $p^*_a$, $p^*_b$, and $d_1$ for the
    imbalance residual model across different values of $N$. Note that $p^*_a$
    and $p^*_b$ here do not coincide with those from the rotation model.}
    \label{tab:imbalance_params}
    \begin{tabular}{@{}lccccc@{}}
        \toprule
        \textbf{N} & \textbf{a} & \textbf{b} & \textbf{$p^*_a$} & \textbf{$p^*_b$} & \textbf{$d_1$} \\
        \midrule
        8{,}000   & $1.360\!\times\!10^{-1}$ & $3.050\!\times\!10^{-1}$ & $1.561\!\times\!10^{-1}$ & $7.816\!\times\!10^{-1}$ & $-1.463\!\times\!10^{0}$ \\
        16{,}000  & $1.551\!\times\!10^{-1}$ & $3.416\!\times\!10^{-1}$ & $1.788\!\times\!10^{-1}$ & $7.949\!\times\!10^{-1}$ & $-1.489\!\times\!10^{0}$ \\
        32{,}000  & $1.497\!\times\!10^{-1}$ & $3.411\!\times\!10^{-1}$ & $1.568\!\times\!10^{-1}$ & $7.767\!\times\!10^{-1}$ & $-1.472\!\times\!10^{0}$ \\
        64{,}000  & $1.530\!\times\!10^{-1}$ & $3.386\!\times\!10^{-1}$ & $1.559\!\times\!10^{-1}$ & $7.689\!\times\!10^{-1}$ & $-1.464\!\times\!10^{0}$ \\
        128{,}000 & $1.320\!\times\!10^{-1}$ & $2.981\!\times\!10^{-1}$ & $1.615\!\times\!10^{-1}$ & $7.782\!\times\!10^{-1}$ & $-1.472\!\times\!10^{0}$ \\
        256{,}000 & $1.337\!\times\!10^{-1}$ & $2.997\!\times\!10^{-1}$ & $1.597\!\times\!10^{-1}$ & $7.768\!\times\!10^{-1}$ & $-1.474\!\times\!10^{0}$ \\
        512{,}000 & $1.383\!\times\!10^{-1}$ & $3.081\!\times\!10^{-1}$ & $1.594\!\times\!10^{-1}$ & $7.809\!\times\!10^{-1}$ & $-1.475\!\times\!10^{0}$ \\
        \bottomrule
    \end{tabular}
\end{table}

The $d_1$ term is close to the value seen in the rotations/$N$ residual fit.
The exact zeroes, however, are not the same as in the rotation model;
there is a small but structured shift. The $p^*_a$ and $p^*_b$ values in
this section therefore do not coincide exactly with those from the
rotation model.

Excluding the $N = 16\text{k}$ case as a small-size outlier, the deviation is
close to constant. Taken together, the two empirical models can be
written as $\text{rot}/N = f(p) + R(p)$ and
$\text{imbalances} \times p = m \times \text{rotations}_{\text{raw}}
+ \lambda \times N \times p + R_h(N, p)$.
There is no mathematical reason these residuals must share a similar
form, yet empirically they do. For that reason, the repeated
warped-cubic structure is best described as an interesting recurring
pattern in the p-AVL dynamics rather than as a final theoretical claim.
As expected, the scalar $K$ also scales with $N$, which sets the size of
the residual.

Notably, the $p^*_a$ and $p^*_b$ values here do not match those from the
rotation model, but instead show a structured deviation. Since $a$ and $b$
are zeroes in $\varphi(f(p))$ space rather than directly in $p$-space, this
shift should be interpreted cautiously, though it is still structured
enough to be noteworthy.

\begin{table}[H]
    \centering
    \caption{Structured deviation $\Delta p^*_a$ and $\Delta p^*_b$ between the
    crossing points of the imbalance residual model and the rotation model,
    across different values of $N$.}
    \label{tab:delta_crossing}
    \begin{tabular}{@{}lcc@{}}
        \toprule
        \textbf{N} & \textbf{$\Delta p^*_a$} & \textbf{$\Delta p^*_b$} \\
        \midrule
        8{,}000   & $+0.035$ & $-0.086$ \\
        16{,}000  & $+0.016$ & $-0.104$ \\
        32{,}000  & $+0.038$ & $-0.085$ \\
        64{,}000  & $+0.040$ & $-0.083$ \\
        128{,}000 & $+0.036$ & $-0.092$ \\
        256{,}000 & $+0.038$ & $-0.091$ \\
        512{,}000 & $+0.038$ & $-0.093$ \\
        \bottomrule
    \end{tabular}
\end{table}

% -------------------------------------------------------
\section{Height, Average Depth, and Pareto Tradeoff}

One would expect the regime $0 < p < 1$ to interpolate monotonically
between BST-like and AVL-like behaviour, with very small $p$ still
allowing longer chains and larger variability. The experiments support
that broad picture. At the same time, insertions along a long path
create roughly $p \times \text{path-length}$ opportunities for repair, which
makes unchecked growth statistically unlikely even when $p$ is small,
though formalising that argument is difficult.

Even a small value of $p$ causes a rapid collapse of the height away from
the BST regime, while the growth remains logarithmic. Average depth
behaves similarly, although its collapse is less trivial to model
cleanly. Many fitting families were tried at the exploratory stage; here
the important point is not the exact fit family, but the clear and
smooth structural transition as $p$ increases.

The collapse is approximately consistent with a $c(p)/\log N$ type scaling,consistent with classical results on BST height [2],
although finite-size effects and small self-intersections prevent
treating that as a final law. The main point is simpler: height and
average depth remain broadly logarithmic. Extremely imbalanced trees are
still possible under probabilistic balancing, but they become
increasingly rare because later insertions create more chances for
earlier imbalances to be repaired, and because a single well-placed
rotation can remove a large amount of structural stress.

However, curve fitting of average depth, which has a similar structure
to average height, resists a trivial collapse with $\log N$ or
$\log N \cdot p$. Several families were tried for the individual $N$, $p$
curves, including Mittag-Leffler, Gompertz, rational, Hill, and
power-law-in-Hill forms. Visually, the curve is smooth, appears
analytic, and is strongly bounded at $p = 0$ and $p = 1$.

\begin{figure}[ht]
    \centering
    \includegraphics[width=0.9\textwidth]{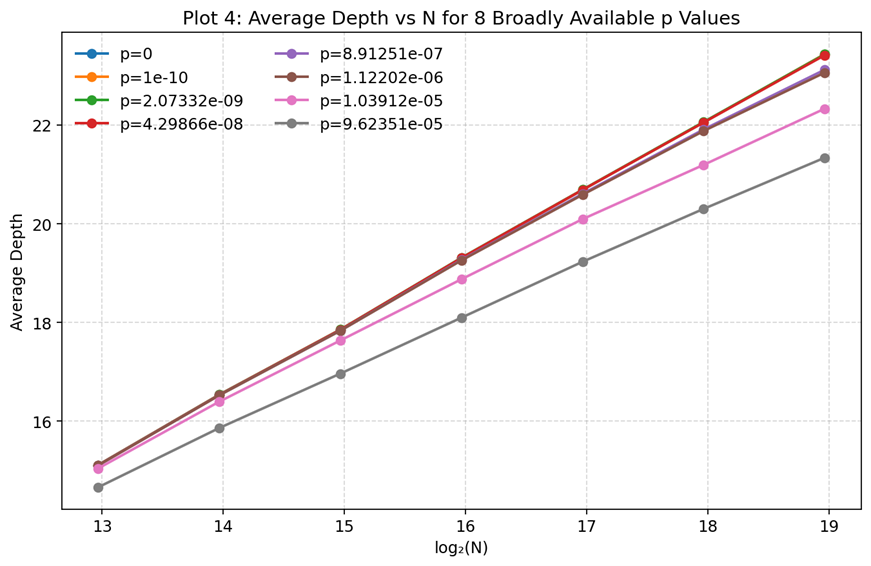}
    \caption{Average depth versus $p$ for multiple values of $N$, using
    logarithmic spacing in $p$ with dense sampling near $10^{-6}$ to $10^{-3}$.
    Average depth changes very rapidly in this small-$p$ region, and the
    variance also drops sharply.}
    \label{fig:avg_depth}
\end{figure}

This plot shows why the code uses logarithmic spacing in $p$ with
especially dense sampling near $10^{-6}$ to $10^{-3}$: average depth changes
very quickly there, and the variance also drops sharply. This naturally
raises the next question: whether the probabilistic relaxation produces
a heavy tail of bad height or depth outliers. The next plots examine
that issue directly. They also examine the difficulty of a trivial
collapse.

\begin{figure}[H]
    \centering
    \includegraphics[width=0.48\textwidth]{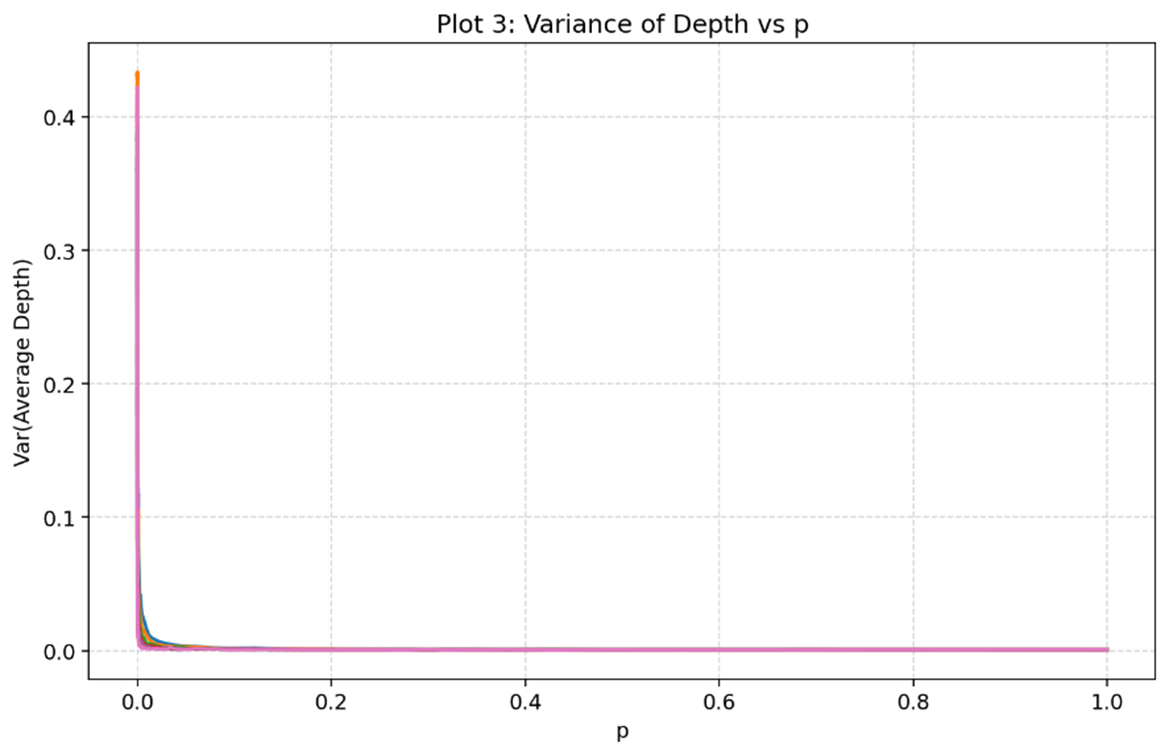}\hfill
    \includegraphics[width=0.48\textwidth]{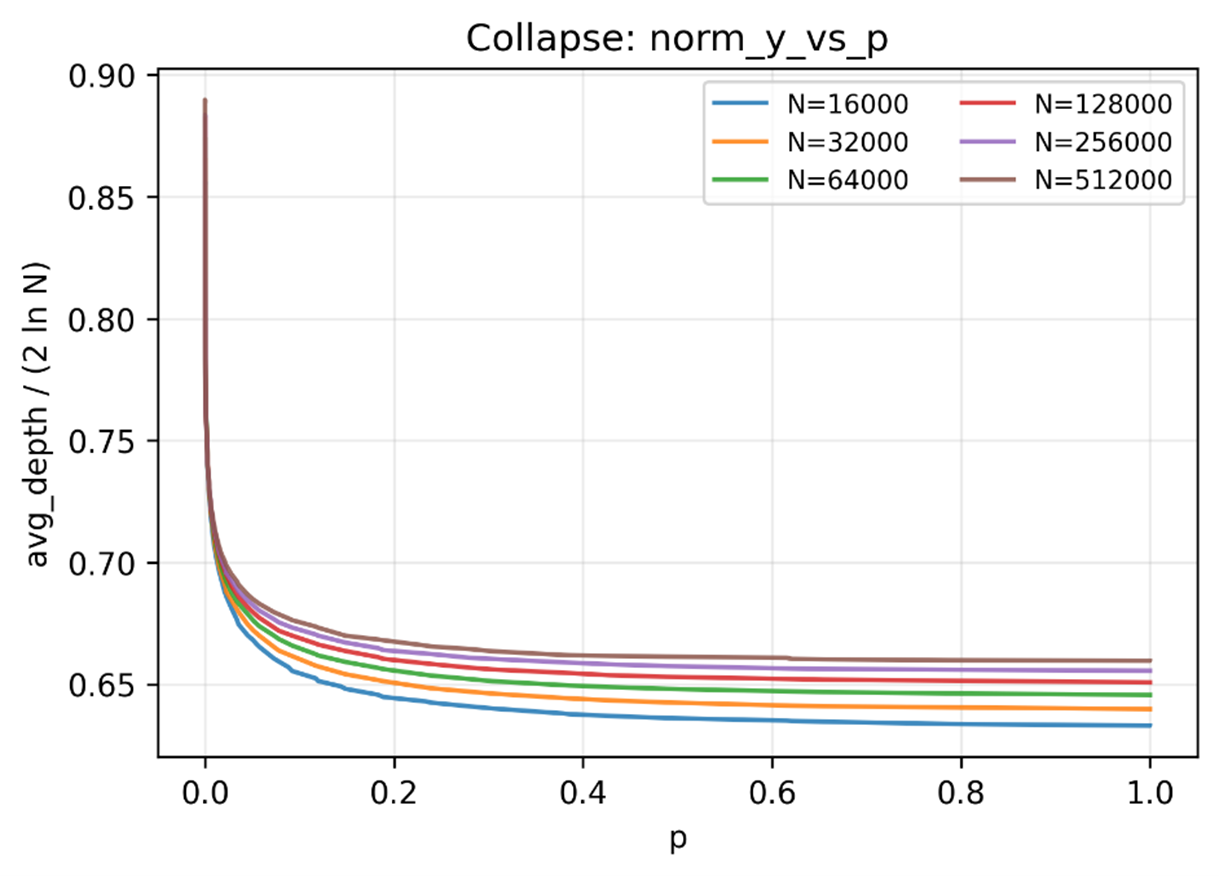}
    \caption{Left: variance of average depth across runs for
    values of $p$, illustrating the rapid collapse of variance as $p$
    increases. Right: rescaled depth curves showing the difficulty of
    achieving a trivial $\log N$ collapse.}
    \label{fig:depth_distribution}
\end{figure}

\begin{figure}[ht]
    \centering
    \includegraphics[width=0.9\textwidth]{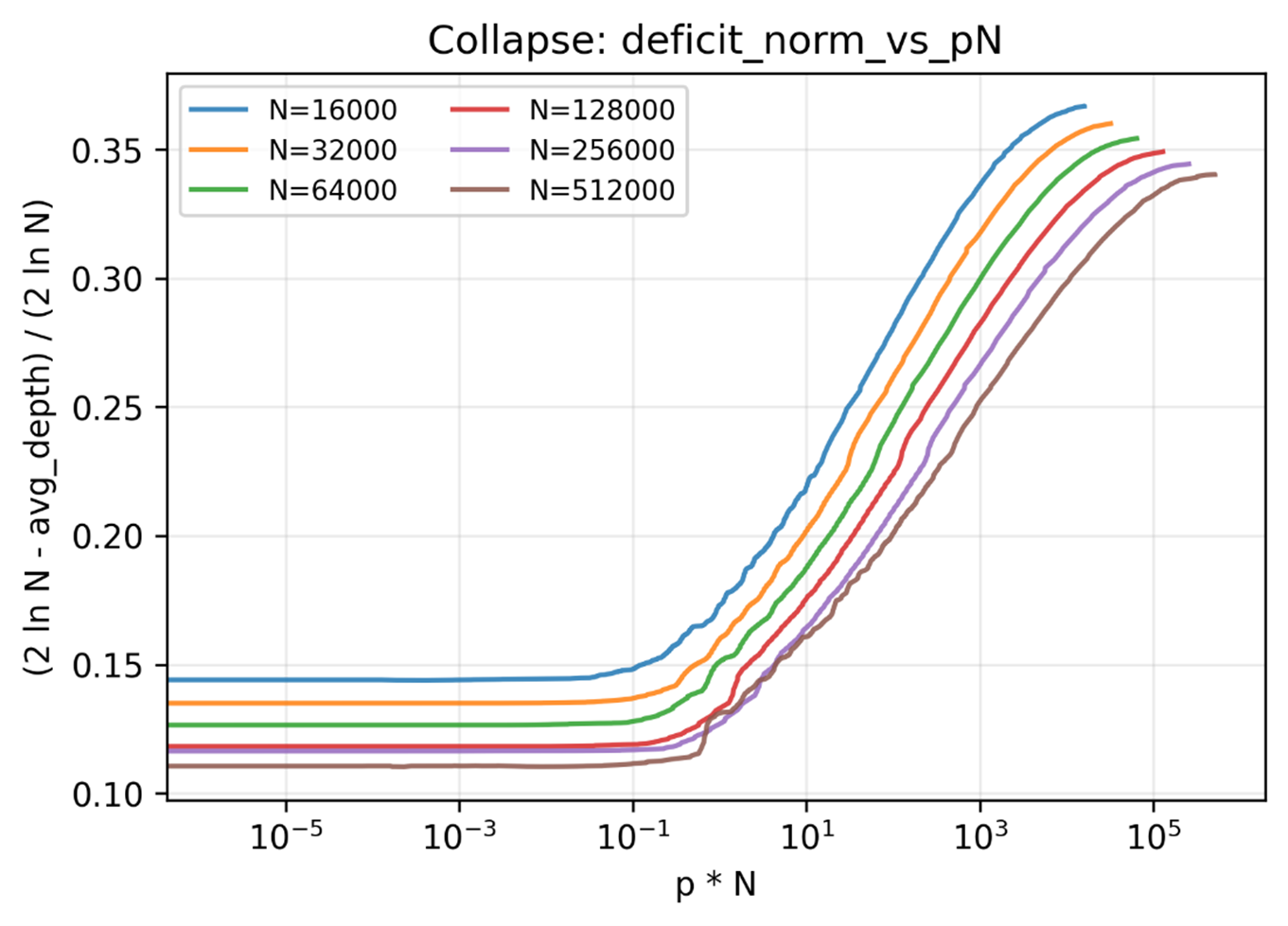}
    \caption{A collapse attempt using $p$ * $N$ ,with a normalized average depth using the BST theoretical value for random insertion, for $N$ from
    8{,}000 to 512{,}000. Finite-size effects and self-intersections prevent a
    clean collapse.}
    \label{fig:depth_logN}
\end{figure}

\begin{figure}[ht]
    \centering
    \includegraphics[width=0.9\textwidth]{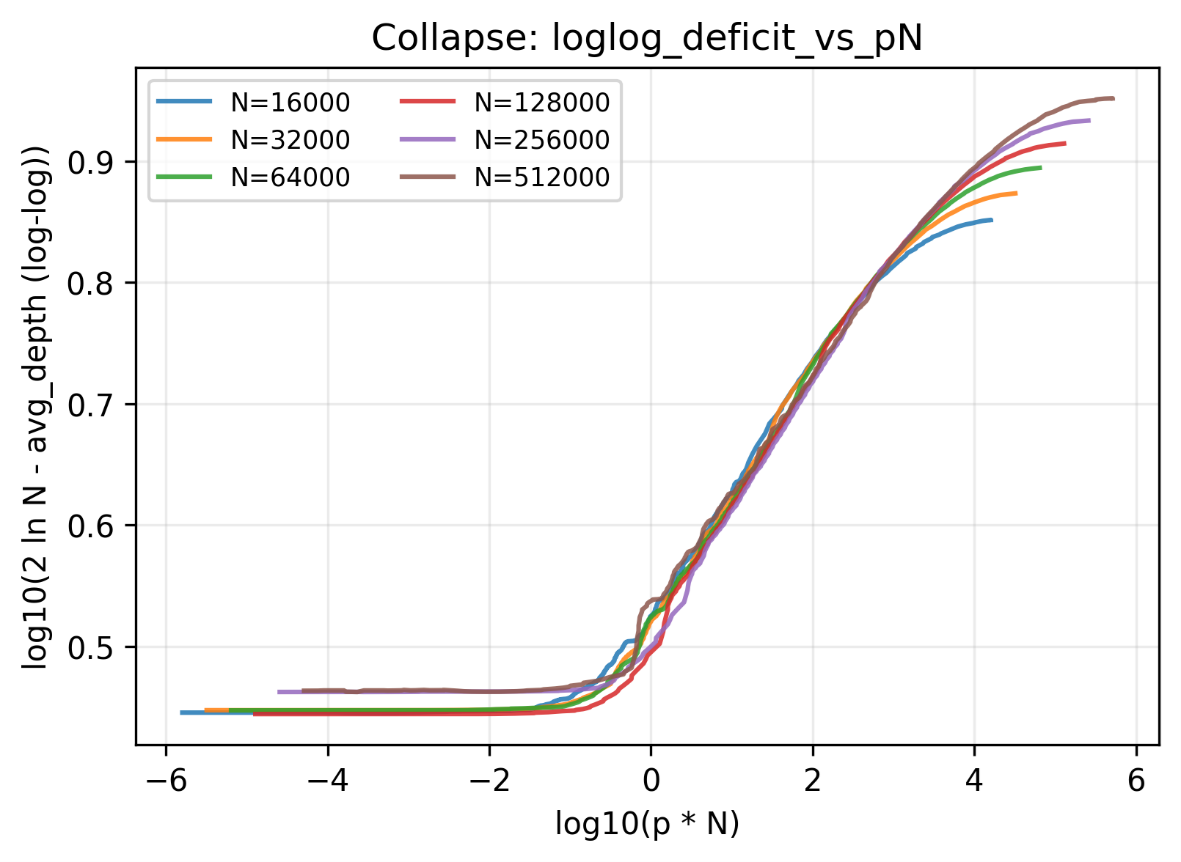}
    \caption{Another unique collapse attempt using a log-log deficit,demonstrating self intersections and the difficulty of a trivial collapse}
    \label{fig:avg_height}
\end{figure}

Each chosen $p$-value was averaged over more than 200 runs on average,
although resolving the far tail more sharply would require even more
sampling. Since $p$ is continuous on $[0,1]$, the state space of possible
trees is very large, so pathological outcomes can in principle still
occur even when they are rare. This also motivates the Pareto tradeoff between balancing cost, measured
by rotations/$N$, and search-related gain, modelled here by average depth.
The experiments use random-order insertion. Even $p = 0.99$ could, in principle, miss every repair opportunity, while
even $p = 0.000001$ could happen to repair enough early imbalances to
create a very balanced tree. Pathological outliers remain possible, but
the evidence suggests they become extremely rare. This motivates looking
for a metric whose distribution may stabilise for fixed $p$ as $N$ grows.
That metric is $\sigma$, which also captures structural imbalance directly. We now turn to the Pareto trade-off, a core part of the paper. The
question is whether the probabilistic relaxation in balancing cost,
measured by rotations, yields a favourable gain in expected search cost,
modelled here by average depth.

\begin{figure}[ht]
    \centering
    \includegraphics[width=0.7\textwidth]{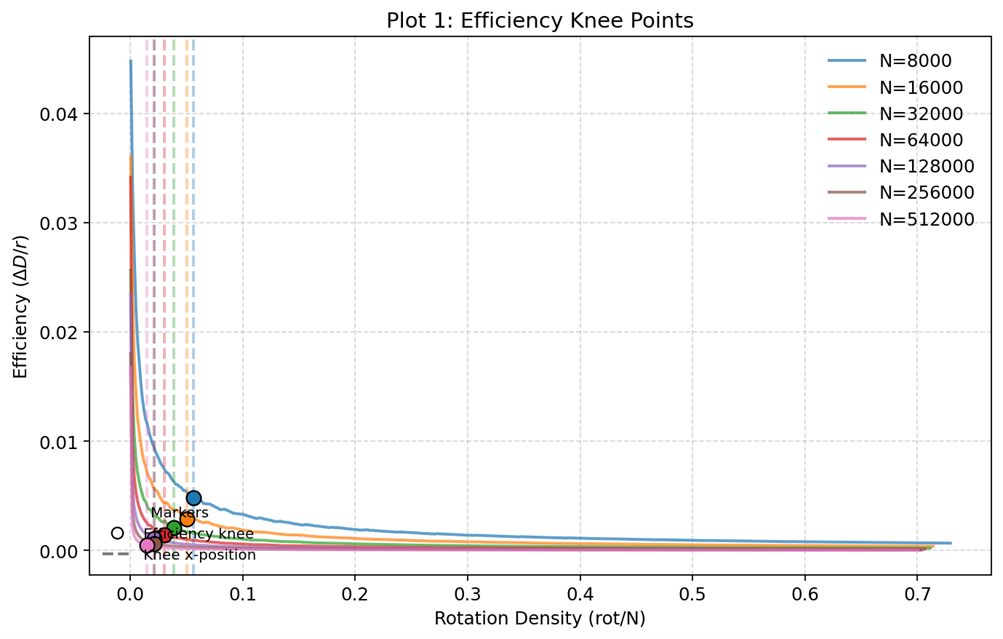}
    \caption{Efficiency curve derived from the Pareto frontier: marginal average
    depth improvement per unit rotation cost
    ($\Delta\text{depth} / \Delta\text{rotations}$), showing that efficiency
    knees appear much earlier in $p$-space than raw Pareto knees.}
    \label{fig:pareto1}
\end{figure}

\begin{figure}[ht]
    \centering
    \includegraphics[width=\textwidth]{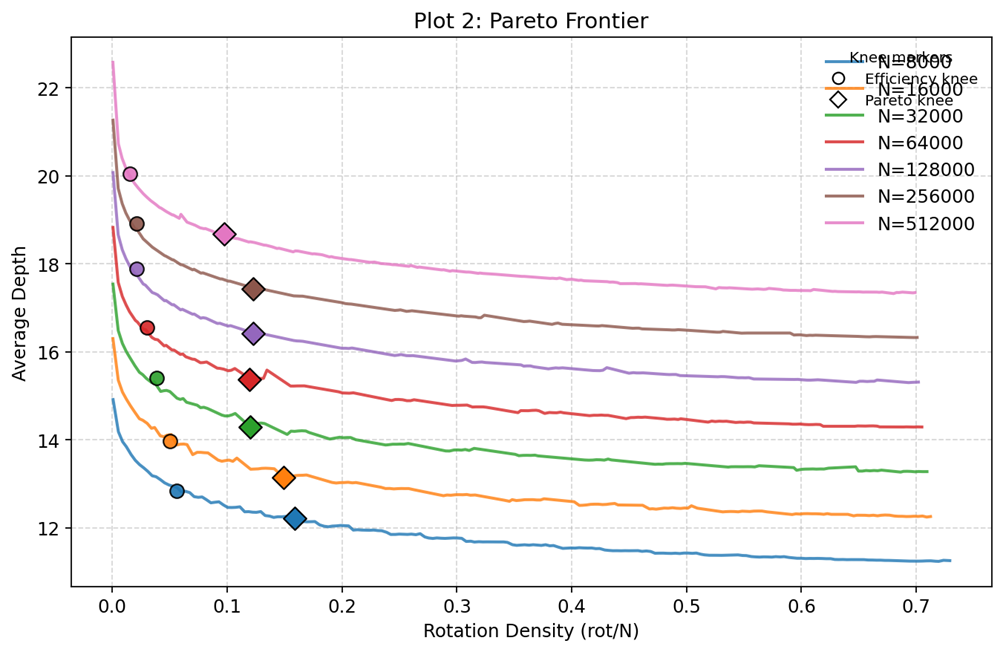}
    \caption{Pareto frontier between rotation cost and average depth gain.
    Each point corresponds to a value of $p$, showing the trade-off between
    balancing effort and expected search efficiency.}
    \label{fig:pareto2}
\end{figure}

We did not explicitly model other insertion orders in detail, though a
small auxiliary experiment suggests the same qualitative picture still
holds: small $p$ already causes a rapid change in average depth, even if
the raw rotation curve differs under more adversarial input orders.
(Around $p=0.01$ for $N=16{,}000$ an adversarial average depth of 45 was
observed, but at just $p=0.08$ it dropped to 19.)

The raw Pareto frontier is shown below. Binning with an intermediate
smoothing value was used to obtain readable curves. The highlighted
points mark two different notions of a knee: one from the efficiency
curve, based on $\Delta\text{depth} / \Delta\text{rotations}$, and one from the
raw cost-gain frontier itself.

The efficiency knees appear much earlier in $p$-space than the raw Pareto
knees. This means that a relatively small amount of balancing already
captures a large share of the structural gain before the rotation cost
becomes comparable to later improvements. Below are the raw statistics.

\begin{table}[ht]
    \centering
    \caption{Efficiency and raw Pareto knee statistics across $N$. The
    efficiency knee (based on $\Delta\text{depth}/\Delta\text{rotations}$)
    appears at much smaller $p$ than the raw Pareto knee, indicating early
    saturation of structural gain.}
    \label{tab:pareto_knees}
    \resizebox{\textwidth}{!}{%
    \begin{tabular}{@{}lcccccc@{}}
        \toprule
        \textbf{N} & \textbf{knee\_p} & \textbf{knee\_rot} & \textbf{knee\_depth} &
        \textbf{pareto\_p} & \textbf{pareto\_rot} & \textbf{pareto\_depth} \\
        \midrule
        8{,}000   & 0.0058 & $0.081\times$ & $1.141\times$ & 0.0298 & $0.228\times$ & $1.086\times$ \\
        16{,}000  & 0.0047 & $0.072\times$ & $1.139\times$ & 0.0280 & $0.214\times$ & $1.072\times$ \\
        32{,}000  & 0.0033 & $0.055\times$ & $1.161\times$ & 0.0196 & $0.172\times$ & $1.076\times$ \\
        64{,}000  & 0.0025 & $0.043\times$ & $1.158\times$ & 0.0184 & $0.171\times$ & $1.075\times$ \\
        128{,}000 & 0.0017 & $0.031\times$ & $1.168\times$ & 0.0184 & $0.175\times$ & $1.072\times$ \\
        256{,}000 & 0.0017 & $0.031\times$ & $1.159\times$ & 0.0184 & $0.176\times$ & $1.068\times$ \\
        512{,}000 & 0.0011 & $0.022\times$ & $1.156\times$ & 0.0131 & $0.140\times$ & $1.077\times$ \\
        \bottomrule
    \end{tabular}}
\end{table}

% \begin{figure}[ht]
%     \centering
%     \includegraphics[width=\textwidth]{figures/image16.png}
%     \caption{Raw Pareto frontier (rotation cost vs.\ average depth), with
%     efficiency and raw knee points highlighted. Binned and smoothed curves
%     are shown for readability.}
%     \label{fig:pareto_frontier}
% \end{figure}
% \vspace{-9cm}
A small addition here for height and the related plot. Height can be
seen to drop rapidly and retain a shape similar to average depth; the
plot below demonstrates the probability that a tree exceeds the expected
height by a certain amount, showing a rapid collapse also.

\begin{figure}[ht]
    \centering
    \includegraphics[width=\textwidth]{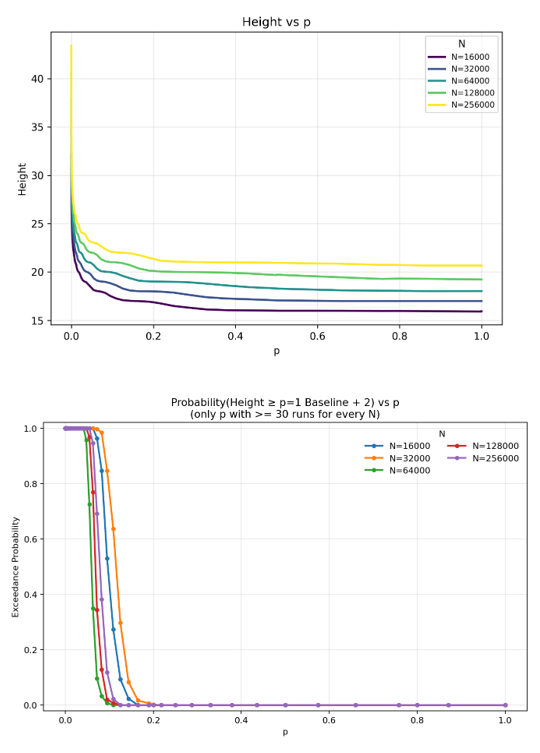}
    \caption{Exceedance probability of tree height: probability that a tree
    exceeds the expected height by a given margin, for several values of $p$.
    The rapid collapse as $p$ increases confirms that extreme height events
    become very rare even at small nonzero $p$.}
    \label{fig:height_exceedance}
\end{figure}
\FloatBarrier
% -------------------------------------------------------
\section{Sigma}
% \vspace{-2cm}
Formally, $\sigma = \frac{1}{N}\sum_{\text{nodes}} \max(0,\,|BF|-1) \times \text{subtree\_size}$.
For $p = 1$, $\sigma = 0$ for an AVL tree; for $p = 0$ it is much larger in
the BST-like regime. Intermediate $p$ gives intermediate $\sigma$. This
statistic is intended as a weighted global imbalance measure:
\textit{violating\_fraction} only counts the final number of violating nodes,
while $\sigma$ also weights how much of the tree each imbalance can
influence.

First, let us look at some standard plots involving $\sigma$. $\sigma / \log N$ nearly collapses but retains
finite size effects and some self-intersections.

\begin{figure}[H]
    \centering
    \includegraphics[width=0.70\textwidth]{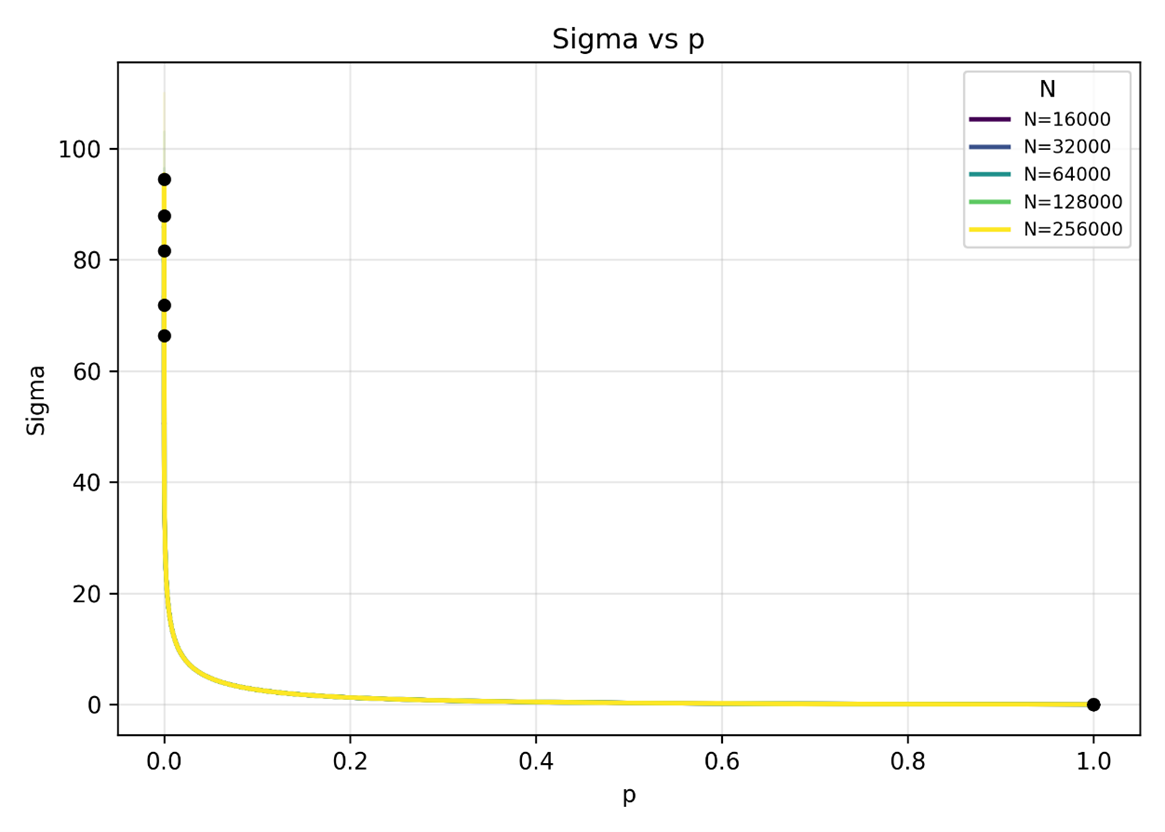}\hfill
    \includegraphics[width=0.70\textwidth]{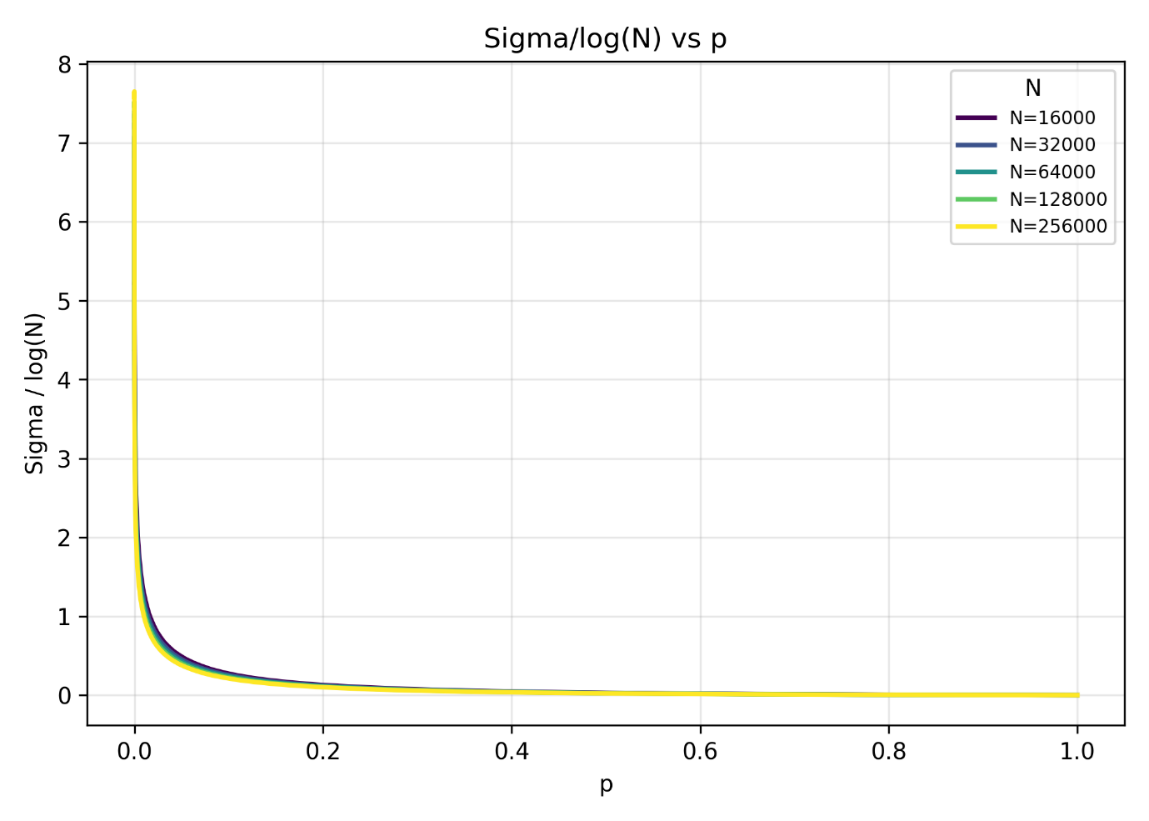}
    \caption{Left: $\sigma$ versus $p$ for multiple $N$, showing
    near-collapse with some dependence on N ,the black dots indicate the value(larger values for larger N)
    Right:
    Demonstration of a near collapse attempt of sigma,indicating sigma/log N is approximately a f(p) but for small p self intersections and imperfect collapse are observed}
    \label{fig:sigma_basic}
\end{figure}

$\sigma$ is useful not only numerically but also conceptually. It acts as a
global structural imbalance measure weighted by subtree influence. There
is a caveat: weighting by subtree size may undervalue some lower nodes
too strongly, so alternative weightings such as $\sqrt{\text{subtree size}}$ or
height-normalised weights could be worth studying later. Even so, $\sigma$
already shows a clear and structured empirical distribution.

At higher $p$ the collapse becomes more discrete, with larger $N$
concentrating into smaller regimes. This is intuitively reasonable: for
fixed $p$, a larger tree creates more opportunities for imbalance
resolution along longer paths.

\begin{figure}[ht]
    \centering
    \includegraphics[width=0.9\textwidth]{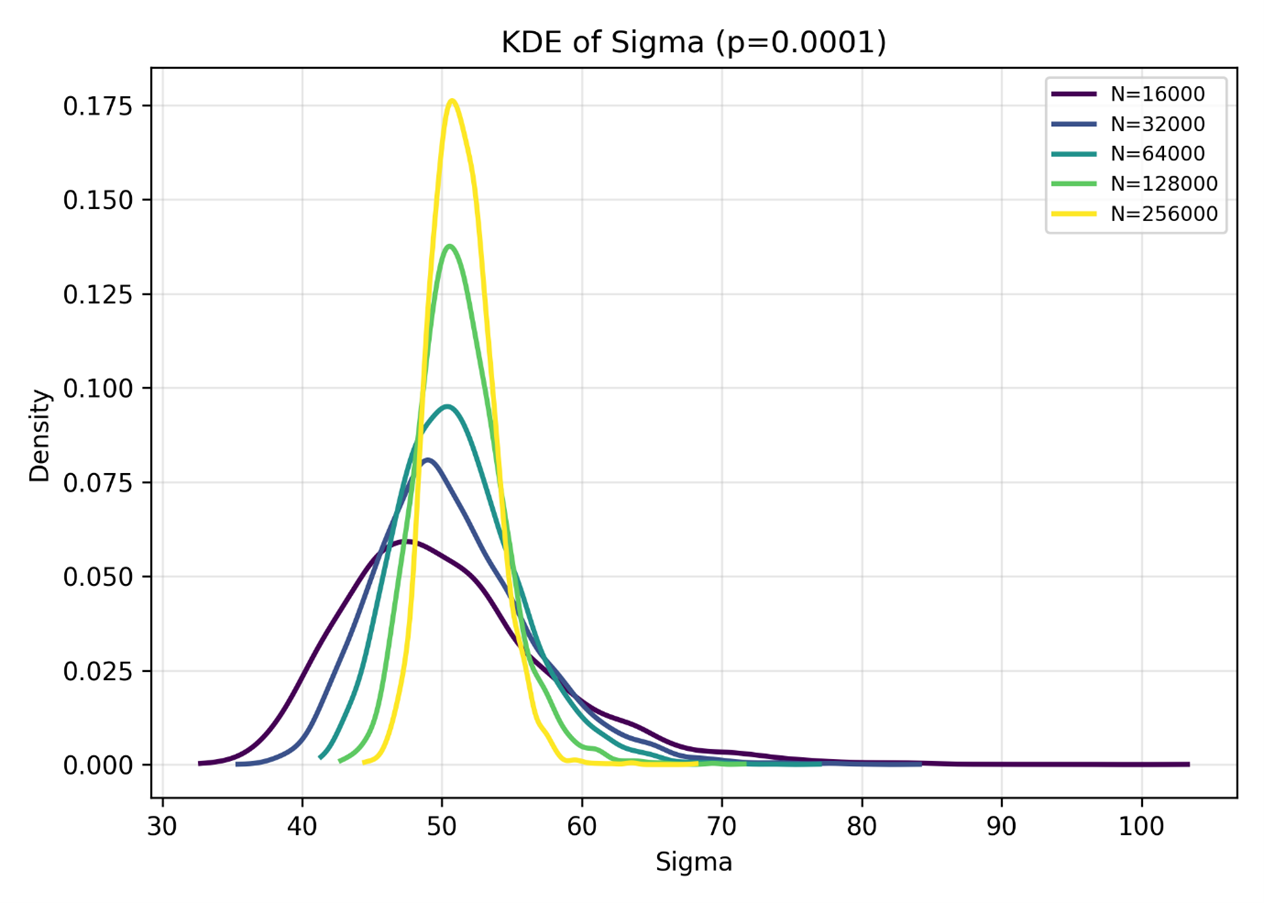}
    \caption{Empirical distribution of $\sigma$ for the stated value  of $p$,
    showing how the distribution tightens and shifts towards a mean value as $p$
    increases. progressive concentration of the distribution as $N$ increases.This is more visible for larger p as compared to the below diagram}
    \label{fig:sigma_dist}
\end{figure}

To support this further, we examine the empirical CDFs below and the
combined ECDF.

\begin{figure}[H]
    \centering
    \includegraphics[width=0.7\textwidth]{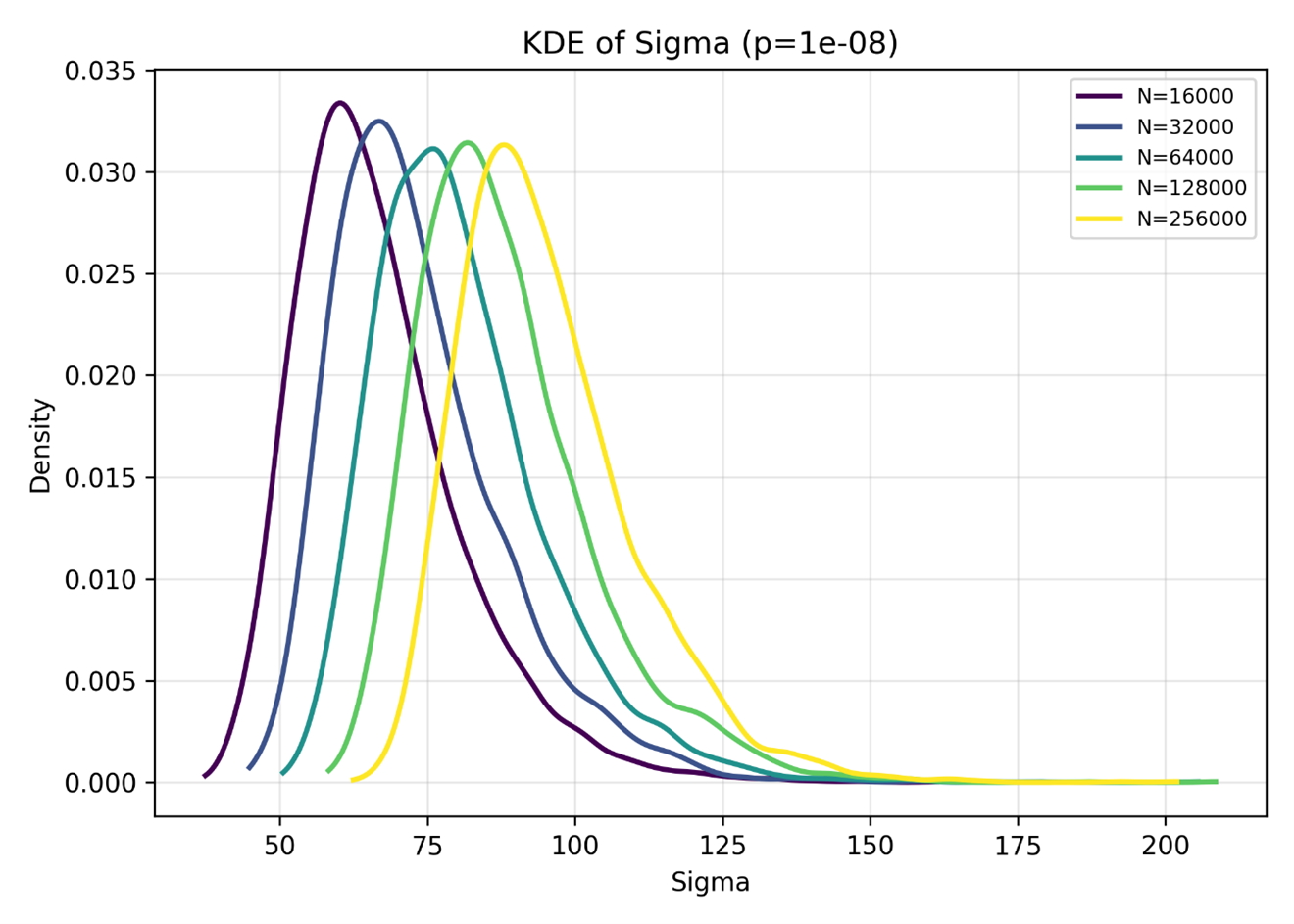}
    \caption{Empirical CDF of $\sigma$ for the given value of  $p$, showing the concentration(minimal at very small p but more pronounced in the above and below plots) as $N$ increases
    }
    \label{fig:sigma_ecdf1}
\end{figure}

\begin{figure}[H]
    \centering
    \includegraphics[width=0.7\textwidth]{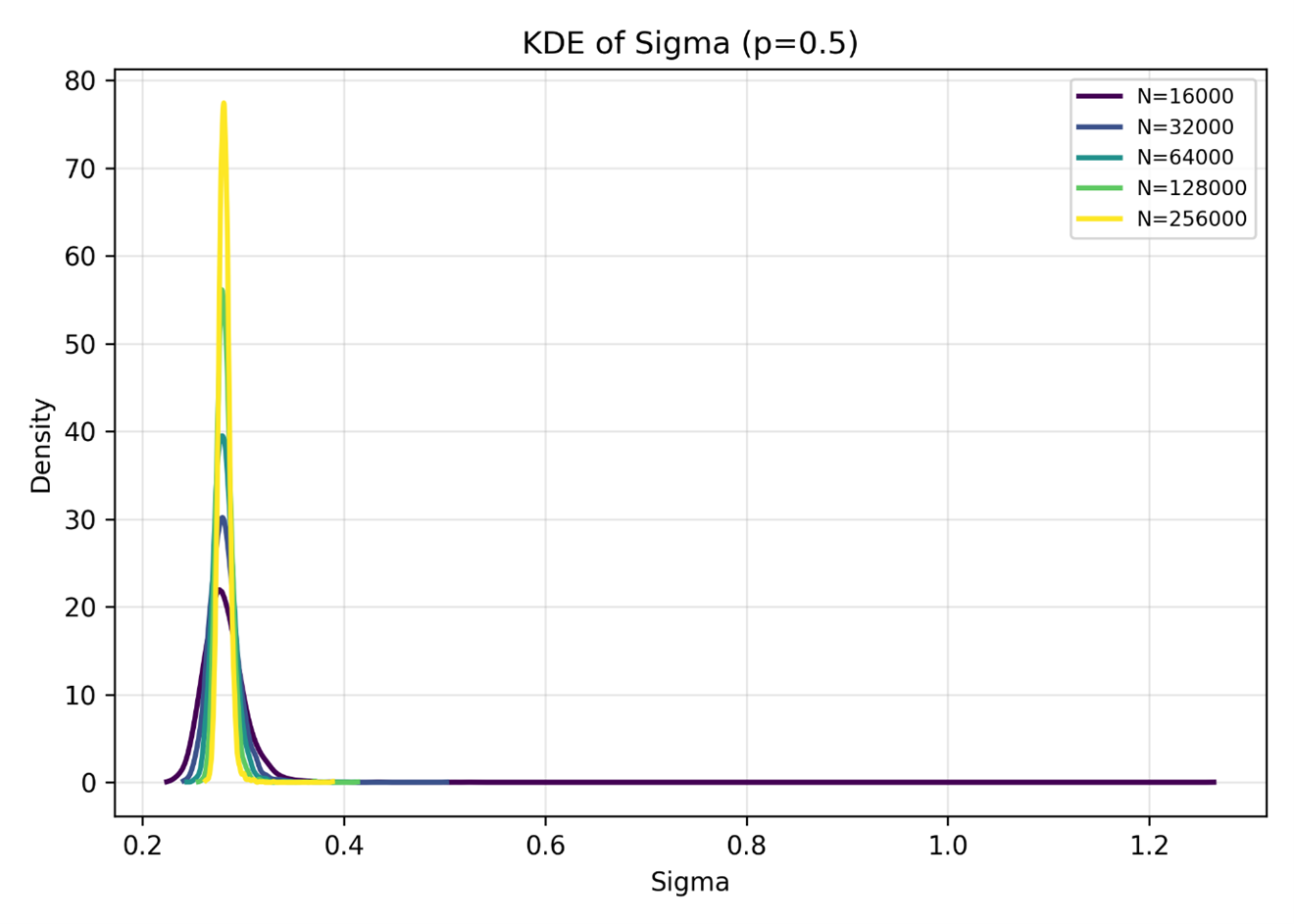}
    \caption{Empirical CDF of $\sigma$ for the given value of  $p$, showing the concentration.Here,at this p(large for p-AVL),a rapid collapse towards the mean value of sigma is seen as $N$ increases}
    \label{fig:sigma_ecdf2}
\end{figure}

\begin{figure}[ht]
    \centering
    \includegraphics[width=0.7\textwidth]{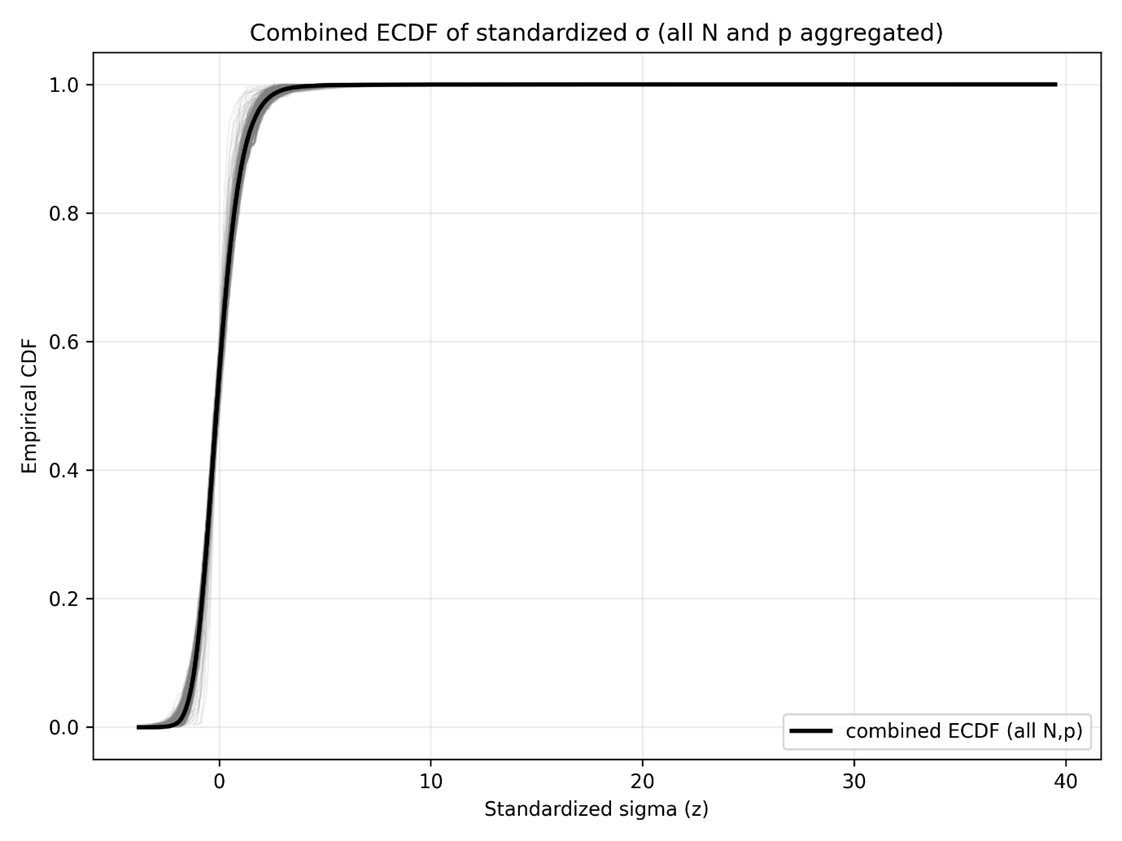}
    \caption{a combined ECDF plot of $\sigma$ highlighting tail behaviour for
    different $p$ values, providing further evidence of distributional
    stability.}
    \label{fig:sigma_ecdf3}
\end{figure}

To further test the similarity of the distribution, we examine tail
probabilities. Ignoring the issue of finite runs, this kind of analysis
could later help in studying probabilistic bounds for extreme structural
events such as unusually large height. For the selected $p$-values shown
here, using more than 4{,}000 runs, the intermediate tail appears close to
exponential.

\begin{figure}[H]
    \centering
    \includegraphics[width=0.7\textwidth]{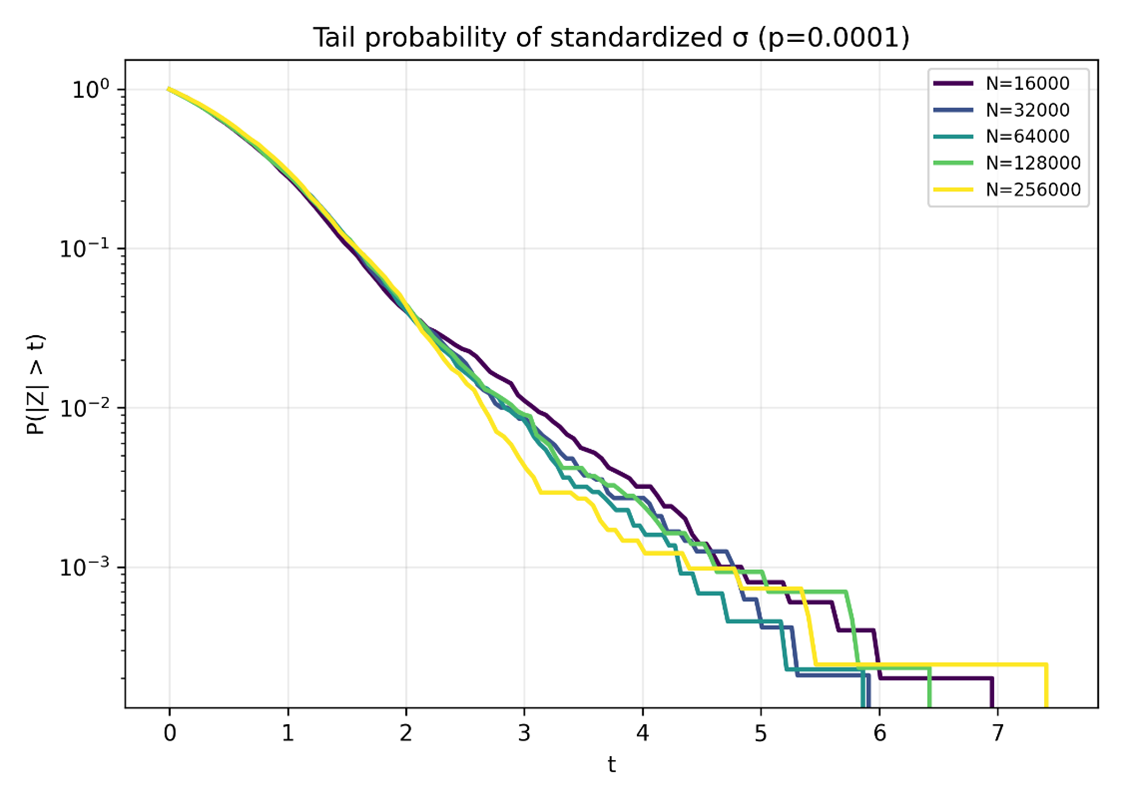}
    \caption{Tail probability of $\sigma$ on a log scale for the stated $p$ value, using more than 4{,}000 runs. The near-linear log-scale
    plots indicate that the intermediate tail is approximately exponential.}
    \label{fig:sigma_tail1}
\end{figure}

\begin{figure}[H]
    \centering
    \includegraphics[width=0.7\textwidth]{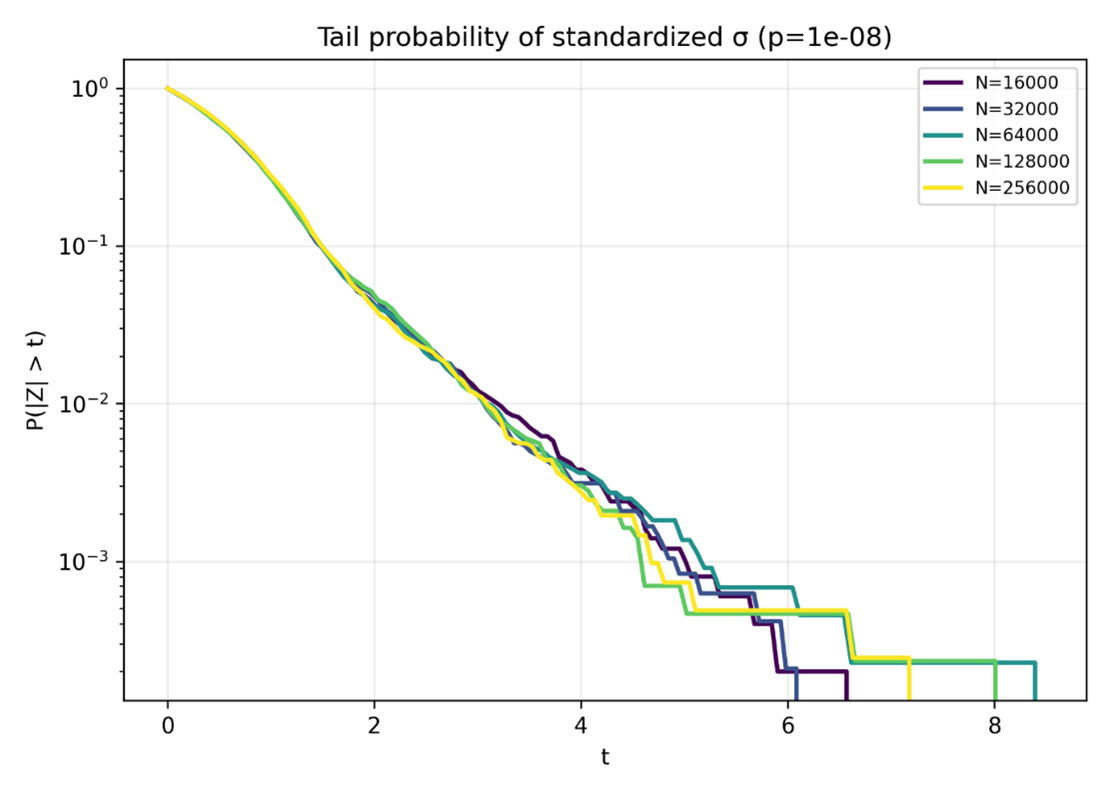}
    \caption{Tail probability of $\sigma$ on a log scale for the stated $p$ value, using more than 4{,}000 runs. The near-linear log-scale
    plots indicate that the intermediate tail is approximately exponential.}
    \label{fig:sigma_tail2}
\end{figure}

\begin{figure}[H]
    \centering
    \includegraphics[width=0.7\textwidth]{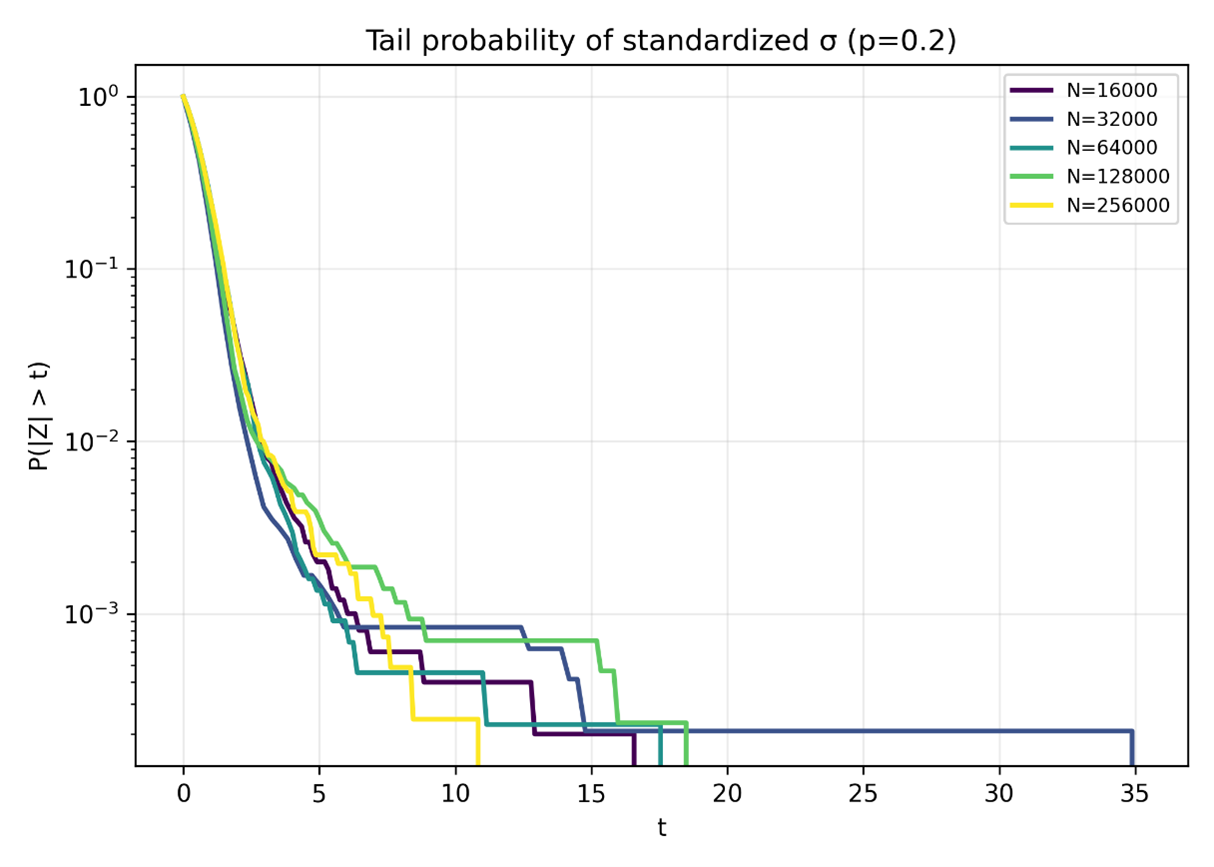}
    \caption{Tail probability of $\sigma$ on a log scale for the stated $p$ value
    , using more than 4{,}000 runs. The near-linear log-scale
    plots indicate that the intermediate tail is approximately exponential.The 3 tail plots indicate across over 8 orders of magnitude the intermediate bounds are approximately exponential}
    \label{fig:sigma_tail3}
\end{figure}

% -------------------------------------------------------
\section*{Conclusion}

In conclusion, we analysed the empirical forms of rotations/$N$, total
imbalances, and the structural statistics of average depth and height,
together with the continuous Pareto frontier in both efficiency and raw
cost form.

The p-AVL rule is simple---probabilistically rebalance each triggered
AVL repair with probability $p$---but the resulting behaviour is not
simple. It provides an exact continuous interpolation from a BST-like
regime to the AVL tree, connecting two well-understood trees through a
single parameter. Rather than giving a single fixed trade-off, it
produces a full Pareto curve from which one may select a preferred
operating regime.

The strongest empirical result is that even small nonzero $p$ already
causes a large structural change. This does not suggest a true phase
transition in the strict sense, but rather a rapid crossover. Since $p$ is
continuous, and since even very small nonzero $p$ can in principle produce
depth-neutral or depth-improving rotations, the apparent picture of `no
change until some critical $p$' should be treated with caution.

Several residual and distributional patterns recur across sections.
These should be treated as promising empirical structure rather than as
final theory, but they suggest that the p-AVL dynamics contain
regularities beyond a trivial interpolation.

From a practical and engineering viewpoint, plain p-AVL may also be seen
as the simplest member of a broader family of schemes in which the
balancing probability depends on some metric of the current tree state.
The most immediate example would be height-based control, where a lower
balancing regime is used normally and a higher one is triggered once the

An important point about the implementation is that compared to the deterministic AVL,which has 2 methods of insertion,top-down and bottom-up.In the p-AVL case,these 2 are not equal,take the simple insertion order of 3->2->1.
a bottom-up update tests the root after the final violation has formed, so it is repaired with probability $p$
A descent-time top-down variant would test the same node before that violation exists, and so need not produce the same law. The results here therefore correspond specifically to the bottom-up formulation used in the experiments.
tree degrades beyond a chosen threshold. In that sense, p-AVL is not
only a standalone model, but also a simple starting point for adaptive
probabilistic balancing rules in the form of $p(M(T))$.where M(T) is a metric of the current tree state.

% -------------------------------------------------------

\end{document}